# A Machine Learning Potential for Graphene

Patrick Rowe

*Thomas Young Centre, London Centre for Nanotechnology, and Department of Physics and Astronomy,*
*University College London, Gower Street, London, WC1E 6BT, U.K.*

Gábor Csányi

*Engineering Laboratory, University of Cambridge, Trumpington Street, Cambridge CB2 1PZ, U.K.*

Dario Alfè

*Thomas Young Centre, London Centre for Nanotechnology and Department of Earth Sciences,*
*University College London, Gower Street, London WC1E 6BT, U.K.*

Angelos Michaelides

*Thomas Young Centre, London Centre for Nanotechnology, and Department of Physics and Astronomy,*
*University College London, Gower Street, London, WC1E 6BT, UK*

(Dated: October 3, 2017)

We present an accurate interatomic potential for graphene, constructed using the Gaussian Approximation Potential (GAP) machine learning methodology. This GAP model obtains a faithful representation of a density functional theory (DFT) potential energy surface, facilitating highly accurate (approaching the accuracy of *ab initio* methods) molecular dynamics simulations. This is achieved at a computational cost which is orders of magnitude lower than that of comparable calculations which directly invoke electronic structure methods. We evaluate the accuracy of our machine learning model alongside that of a number of popular empirical and bond-order potentials, using both experimental and *ab initio* data as references. We find that whilst significant discrepancies exist between the empirical interatomic potentials and the reference data - and amongst the empirical potentials themselves - the machine learning model introduced here provides exemplary performance in all of the tested areas. The calculated properties include: graphene phonon dispersion curves at 0 K (which we predict with sub-meV accuracy), phonon spectra at finite temperature, in-plane thermal expansion up to 2500 K as compared to NPT *ab initio* molecular dynamics simulations and a comparison of the thermally induced dispersion of graphene Raman bands to experimental observations. We have made our potential freely available online at [http://www.libatoms.org].

## I. INTRODUCTION

As a result of its unique mechanical, electronic and structural properties, graphene has been the subject of extensive investigation since it was first isolated.[1–3] These, combined with its characteristic 2D nature, have resulted in graphene becoming the 'poster child' for materials design in nano-electronic, mechanical and optical research.[4,5] It is the fundamental building block of all sp$^2$ hybridized carbon allotropes; graphene may be rolled to form nanotubes or fullerenes, or stacked to form graphite.[3] These similarities are not merely topological, but also extend to the physical properties of the materials; graphene, graphite and carbon nanotubes share many electronic and vibrational properties for this reason.[6–8] It is concerning, therefore, that despite the vast number of excellent computational and experimental publications focused on elucidating the microscopic origins of graphene's unique properties, existing calculations often draw quantitatively or qualitatively conflicting conclusions. In particular, modern empirical potentials provide disparate results, with conflicting predictions made for fundamental properties such as the coefficient of thermal expansion (CTE), even the sign of which is not reliably predicted.[9–12] There are a great number of interesting phenomena associated with graphene, such as the phonon assisted diffusion of small molecules on the graphene surface,[13] the study of thermal transport[14–16] and the incorporation of nuclear quantum effects into simulations which would benefit greatly from a highly accurate graphene model.[17,18]

Empirical and bond-order potentials have long provided an indispensable tool in facilitating molecular dynamics (MD) studies of carbonaceous materials. The first many-body potential for carbon was published in 1988 by Tersoff, which gained rapid acceptance as research into amorphous and other exciting allotropes of carbon (nanotubes and fullerenes) grew.[19,20] Modification and reparameterization of the Tersoff potential made possible the treatment of hydrocarbons and significantly improved the description of the pure carbon allotropes in the form of the Reactive Empirical Bond-Order potential (REBO).[21] While the REBO potential represented a substantial improvement over the Tersoff potential, neither of these accounted for the effects of dispersion interactions and were inherently short ranged in nature. The AIREBO[22] potential aimed to correct this, by explicitly incorporating long-range interactions into the functional form through the use of switching functions, thereby maintaining effectively the same short-range potential as its predecessor, REBO. The description of the bonding behavior of this potential was further improved upon in 2015 by the incorporation of a Morse pair potential (replacing the Lennard-Jones term in the original) to improve the description of anharmonicity in the bonding terms.[22,23] A fully reparametrized bond-order potential was produced by Los and Fasolino in the form of LCBOP, wherein the short range potential was parametrized to include effective long range corrections.[24]

In addition to these developments in traditionally constructed forcefields, a number of different approaches have emerged which show promise as computational tools. The ReaxFF class of potentials do not represent an iterative improvement upon any of the previously discussed empirical carbon potentials, instead adopting an approach centered around the description of bond dissociation and reactivity.[25] The potential constructs the bond order from the interatomic distance, from which is derived the bond energy. Also included in the functional form are terms to account for van der Waals, Coulombic, and over- and under-coordination energies, the terms of which are fitted to quantities such as atomic charges, bond, angle and torsional energies and heats of formation.[25,26] Density Functional Tight Binding (DFTB) represents yet another approach, it is not an interatomic potential in the traditional sense, rather an electronic method which operates on a tightly constrained set of parameterized wavefunctions. DFTB is based on a second order expansion of the DFT total energy into a distance dependent electronic Hamiltonian and two-body repulsive classical term. The diagonal elements of the Hamiltonian matrix correspond to the atomic (s, p, d) eigenenergies, while the distance dependent off-diagonal elements of the Hamiltonian - the bond energies - are parameterized to DFT and evaluated by interpolation.[27,28]

In recent years, maching learning (ML) methodologies have emerged as an exciting tool within chemical and materials science. Applications have included structure prediction,[29,30] property prediction (including atomization energies, band gaps and nuclear chemical shifts)[31–35] and the development of DFT exchange-correlation functionals.[36–39] The application of machine learning algorithms to the development of interatomic potentials also represents an innovative approach which has recently attracted much attention. ML based approaches to the generation of intermolecular potentials are by their very nature parametrized exclusively to *ab initio* data - but the differences between an ML and a bond-order or empirical potential extend far beyond this. The general ML approach makes very different use of *ab intio* data than an empirical many-body potential. While potentials such as LCBOP may optimize the parameters of (for example) a Morse style functional form based on a fit to *ab inito* data, such an approach will always be fundamentally limited by the assumption that the two-body part of such an interaction is describable by a specifc closed mathematical form. This assumption - while physically motivated - does not arise from a first principles consideration of the shape of the potential energy surface (PES), but from empirical observations and will therefore incorporate a physical bias, limiting the quality of the resulting potential. ML approaches, however, make no such assumptions about the functional form into which the PES may be decomposed - beyond that it must be a regular function of the atomic coordinates (continuously differentiable) and that interactions become infinitesimal as interatomic distances become very large. Machine learning methodologies have been shown to be capable of the reproduction of arbitrary functions with arbitrarily high accuracy.[40]

The first attempts at modelling the PES in its full dimensionality using ML methods made use of artificial neural networks, in which the PES was for the first time represented as a sum of atomic contributions to the total energy.[41] This approach was able to accurately reproduce the structural and elastic properties of the crystal structures of graphite and diamond and was used to study the mechanism of the phase transition between the two states.[42,43] The first generally applicable potential for carbon that made use of ML methods came in the form of a GAP designed to treat the amorphous phase of carbon.[29,44] This provided excellent agreement with a number of experimental observations on the properties of amorphous carbon, including bulk moduli, radial distribution functions and topological properties such as the number of rings present in amorphous structures of a given density. Early attempts at the generation of ML models were trained using the readily available DFT total energies,[41] however more efficient use of the training data can be made by training a model on the energies, forces and virial stresses obtainable from DFT; there being 3N data points available in the form of atomic forces compared to the single value for energy available from *ab initio* calculations.[45] A more detailed discussion of the features and approaches to the development of ML potentials can be found elsewhere.[46–49]

In this work we use the Gaussian Approximation Potential method[45] to generate an accurate ML interatomic potential for graphene, with the aim of directly comparing the capabilities of modern machine learning methods with those of empirically constructed many-body potentials. We evaluate the quality of the prediction of atomic forces of our GAP model and a number of empirical potentials versus a reference DFT method. We also compare predictions of the finite temperature phonon spectra of graphene with experimental results, where we find excellent agreement. We further compare the predictions of our GAP potential to those from *ab initio* molecular dynamics (AIMD) simulations of the thermal expansion of graphene - a property which has historically been very challenging for interatomic potentials to predict.[12,50–52] We show thereby that for the case of graphene, machine learning potentials have the capability to act as a substitute for direct *ab initio* calculation, at a much reduced cost and only marginally compromised accuracy. This capability will be particularly valuable in instances where accurate descriptions of dynamics are mandated, such as the description of the diffusion of small molecules on the graphene surface[13] and the treatment of nuclear quantum effects via path integral molecular dynamics.[17,18]

The remainder of this paper will be structured as follows; in section II we provide an outline of how the GAP model is constructed, section III outlines how the *ab initio* configurations and training data were generated. Sections IV to VI are concerned with the evaluation and benchmarking of the potential, considering first the force accuracy, followed by the phonon spectra and thermally induced Raman band dispersion, lattice parameters and thermal expansion. We give our conclusions in section VII.

## II. CONSTRUCTION OF A GAUSSIAN APPROXIMATION POTENTIAL

Gaussian Approximation Potentials are the product of the application of the Gaussian kernel regression machine learning methodology to the problem of function interpolation of the Born-Oppenheimer PES.[45,49] The *ab initio* PES is sampled using a database of observations of quantum mechanical (often DFT) atomic forces and total energies on structures representative of the desired regions of phase space to be studied. These data are used to train the GAP model which can be used to accurately interpolate energies and forces between the previously observed reference data points, the resulting prediction can be used to generate MD trajectories; much like an empirical potential. This method circumvents a problem inherent in empirical potentials wherein assumptions must be made about the functional forms into which the PES can be decomposed. No prior supposition is made, for example, that the microscopic interactions between two atoms must be representable by a harmonic, Morse or Lennard-Jones type function. This allows for a faithful and unbiased (so far as any *ab inito* method may be called unbiased) representation of the PES to be built, which may be conveniently evaluated to accurately predict the energies and forces acting on arbitrary configurations within the sampled phase space.

In the quantum mechanical reference dataset used to generate the potential, only the total energies, forces and virial stresses are available. In order to facilitate the simulation of systems of larger sizes than those upon which *ab initio* calculations are feasible, the GAP model total energy is decomposed into a sum of local contributions, computed from kernel functions which represent the similarity between chemical environments. In this work we decompose the total energy function into a sum of two body (2b), three body (3b) and many-body (MB) interactions, which are weighted (in terms of their contribution to the total energy and atomistic forces) based on their respective statistically measured contributions. The mathematical form of these descriptors is discussed below. The largest portion of the energy is described by pairwise interactions, then 3b, then MB contributions, each of which is represented by a distinct descriptor and associated kernel function.[45,46,53] The descriptor is a transformation of the atomic Cartesian coordinates into a rotationally and translationally invariant form which is suitable for use as input to a ML algorithm. Descriptors vary greatly in their complexity, the 2b term used here is simply the distance between two atoms, while the MB term takes the form of the smooth overlap of atomic positions (SOAP) descriptor, which provides an overcomplete mapping of general n-body configurations. There are many other possible descriptors in the literature, including symmetry functions, Coulomb matrices and bispectra.[48,54,55] We choose this combined descriptor machine learning model as it has been previously shown to greatly improve the stability of a GAP model for amorphous carbon.[44] We also found in the development of our potential that combined descriptors additionally facilitated greater accuracy - a higher quality potential - thereby making more efficient use of the training data as compared to single descriptor methods.

The fundamental feature defining an interatomic potential is that the total energy is the sum of individual atomic contributions. The local atomic energy expression for the GAP model is a linear combination over the contributions from each kernel function $K^{(d)}$ associated with a descriptor $d$:

$$\epsilon^{(d)}(\mathbf{q}^{(d)}) = \sum_{t=1}^{N_t^{(d)}} \alpha_t^{(d)} K^{(d)}(\mathbf{q}_i^{(d)}, \mathbf{q}_t^{(d)}), \qquad (1)$$

in which the sum over $t$ runs over the $N_t$ basis functions. $K^{(d)}(\mathbf{q}_i^{(d)}, \mathbf{q}_t^{(d)})$ is the covariance kernel quantifying the similarity between the descriptor of the atomic environment for which the prediction is to be made, $\mathbf{q}_i^{(d)}$, and the prior observation, $\mathbf{q}_t^{(d)}$, which has associated with it a weighting $\alpha_t$ obtained during the fitting process. The total energy expression for a system is then given by the sum of each of the contributions of each descriptor used in the model, weighted by a corresponding factor $\delta$

$$\begin{aligned} E = &\delta^{(2b)} \sum_{ij} \epsilon^{(2b)}(\mathbf{q}_{ij}^{(2b)}) \\ &+ \delta^{(3b)} \sum_{ijk} \epsilon^{(3b)}(\mathbf{q}_{ijk}^{(3b)}) \\ &+ \delta^{(MB)} \sum_{i} \epsilon^{(MB)}(\mathbf{q}_i^{(MB)}). \end{aligned} \qquad (2)$$

The indices $i$, $j$ and $k$ run over all atoms in the system. We now introduce the mathematical form of each of the descriptors used. The two body descriptor is simply the distance between any two atomic pairs $i$ and $j$,

$$q_{ij}^{(2b)} = |\mathbf{r}_j - \mathbf{r}_i| \equiv r_{ij}, \qquad (3)$$

where $\mathbf{r}_j$ indicates the position vector of atom $j$. The 3b term ($\mathbf{q}^{(3b)}$) used here involves a symmetrized transformation of the Cartesian coordinates, which is designed to be permutationally invariant to the swapping of atoms $j$ and $k$, given by[49]

$$\mathbf{q}_{ijk}^{(3b)} = \begin{pmatrix} r_{ij} + r_{ik} \\ (r_{ij} - r_{ik})^2 \\ r_{jk} \end{pmatrix}. \qquad (4)$$

Many body interactions are described using the recently introduced SOAP descriptor.[48,53] For this descriptor, we begin with the atomic neighbor density around an atom $i$, which is constructed by the placement of a Gaussian function on each neighbor atom $j$ within a given cut-off $r_{cut}$,

$$\rho_i(\mathbf{r}) = \sum_j f_{\text{cut}}(r_{ij}) \exp\left[-\frac{(\mathbf{r}_i - \mathbf{r}_{ij})^2}{2\sigma_{\text{at}}^2}\right]. \qquad (5)$$

Here, $\sigma_{\text{at}}$ determines the width of the Gaussian and $f_{\text{cut}}$ is any function which goes smoothly to 0 at the cut off distance (we note that all descriptors in this work use this same cut-off function). For example,

$$f_{\text{cut}}(r_{ij}) = \begin{cases} 1 & \text{if } r_{ij} \leq r_{\text{cut}} - w_{\text{cut}} \\ g_{\text{cut}}(r_{ij}) & \text{if } r_{\text{cut}} - w_{\text{cut}} < r_{ij} \leq r_{\text{cut}} \\ 0 & \text{if } r_{ij} > r_{\text{cut}} \end{cases} \quad (6)$$

in which $w_{\text{cut}}$ specifies the width of the region over which the function goes to 0, and where $g_{\text{cut}}(r_{ij})$ may be any function which decreases monotonically from 1 to 0 between $r_{\text{cut}} - w_{\text{cut}}$ and $r$. We choose

$$g_{\text{cut}}(r_{ij}) = \frac{1}{2}\left[\cos\left(\pi \frac{r_{ij} - r_{\text{cut}} + w_{\text{cut}}}{w_{\text{cut}}}\right) + 1\right]. \quad (7)$$

The neighbor density is then expanded in a basis set of radial functions $g_n(r)$ and spherical harmonics $Y_{lm}(\mathbf{r})$ as

$$\rho_i(\mathbf{r}) = \sum_{nlm} c^{(i)}_{nlm} g_n(r) Y_{lm}(\mathbf{r}), \quad (8)$$

in which $c^{(i)}_{nlm}$ are the expansion coefficients for the atom $i$. The descriptor itself is formed from the power spectrum of these coefficients

$$\mathbf{q}^{\text{MB}}_i = p^{(i)}_{nn'l} = \frac{1}{\sqrt{2l+1}} \sum_m c^{(i)}_{nlm} (c^{(i)}_{n'lm})^*. \quad (9)$$

To obtain a power spectrum for $n < n_{\max}, l < l_{\max}$, the expansion of the atomic neighbor density into radial basis functions can employ a truncated basis set. In the local energy expression (eq. 1) the covariance kernel $K^{(d)}_t$ provides a quantitative measure of the similarity between two chemical environments $\mathbf{q}^{(d)}$ and $\mathbf{q}^{(d)}_t$. The functional form of the covariance kernel differs depending on the descriptor, for the 2b and 3b descriptors, we choose the squared exponential kernel,

$$K^{(d)}(\mathbf{q}^{(d)}_i, \mathbf{q}^{(d)}_t) = \exp\left[-\frac{1}{2} \sum_\xi \frac{(q^{(d)}_{\xi,i} - q^{(d)}_{\xi,t})^2}{\theta^2_\xi}\right]. \quad (10)$$

The index $\xi$ runs over either the single value of the 2b descriptor, or the three components of the 3b descriptor. For the many-body SOAP descriptor, the natural choice of covariance function is the dot product of the two power spectra $\mathbf{p}_i$ and $\mathbf{p}_t$ with elements $p^{(i)}_{nn'l}$ and $p^{(t)}_{nn'l}$, as this corresponds analytically to an integrated overlap over all possible 3D rotations of the two associated neighbor densities, that is

$$K^{(\text{MB})}(\mathbf{q}^{(\text{MB})}_i, \mathbf{q}^{(\text{MB})}_t) = [\mathbf{p}_i \cdot \mathbf{p}_t]^\zeta = \left[\int d\hat{R} \left|\int dr^3 \rho_i(\mathbf{r}) \rho_t(\hat{R}\mathbf{r})\right|^2\right]^\zeta. \quad (11)$$

### III. GENERATION OF TRAINING DATA

Our training data are generated from tightly converged plane-wave DFT calculations performed on configurations sampled from various molecular dynamics trajectories. While the atomic configurations herein are generated using a variety of methods (MD with existing potentials and various iterations of our GAP model) the values for atomic forces, virial stresses and energies which comprise the training dataset have all been calculated using precisely the same level of DFT. For these calculations, we use the VASP plane-wave DFT code,[56–58] with the optB88-vdW dispersion inclusive functional[59,60] with a projector augmented wave potential,[61] a plane wave basis cutoff of 650 eV and Gaussian smearing of 0.05 eV.[62,63] We use a dense reciprocal space Monkhorst-Pack grid[64] with a maximum spacing of 0.012 Å$^{-1}$. In order to ensure a low degree of noise on the calculated forces, the energy convergence criterion for the SCF iterations was set to $10^{-8}$ eV. We choose the optB88-vdW functional as it has already been shown to provide an excellent description of graphitic carbon.[65] We further evaluate the sensitivity of our predictions to this choice, by comparing against other common exchange-correlation functionals, which is discussed briefly in section IV and the details of which are given in the Supplemental Material (SM).[66]

The first set of training data was generated from three MD simulations of a free-standing graphene sheet comprised of 200 atoms with lattice parameter a = 2.465 Å. Simulations were performed in the NVT ensemble at temperatures of 1000, 2000 and 3000 K. Trajectories were generated using the LAMMPS[67] open source molecular dynamics program, interactions were modelled using the LCBOP many-body potential for carbon and a Nosé-Hoover thermostat was used to maintain a constant temperature over the simulation. A total of 100 configurations were sampled from each of the three 2 ns trajectories at 20 ps intervals, the total energies and forces of these atomic configurations were then calculated using VASP as outlined above.

An initial GAP model was generated using the *ab initio* quantities computed on the 300 configurations. A further set of molecular dynamics trajectories were generated as above, but with interactions now computed using the preliminary GAP model. Simulations were performed between 300 and 3000 K at a fixed lattice parameter of a = 2.465 Å, a sample of *ab initio* energies, forces and virial stresses from these configurations was added to the training set to produce a second GAP model. A number of iterations of improvement were performed using this approach, the final dataset was comprised of 1083 configurations of 200 atoms at temperatures between 300 and 4000 K and lattice parameters between 2.460 and 2.480 Å.

A random sample of 5% of these configurations was withheld as a validation set to benchmark the quality of the GAP fitting procedure. The parameters used for the fitting of the GAP model are shown in Table. I. Additionally, we choose the expected error (analogous to the target closeness of the fit of the GAP model to the training data) in energies to be $\sigma_E = 10^{-3}$ eV, for forces we choose $\sigma_f = 5 \times 10^{-4}$ eV, and for virial stresses $\sigma_v = 5 \times 10^{-3}$. The training configurations and GAP model files developed herein are freely available in our online repository at [http://www.libatoms.org]. The QUIP source code, necessary to make use of the GAP model, is also available online at [https://github.com/libAtoms/QUIP].

## IV. FORCE PREDICTION

The first natural metric for the quality of a potential - in particular one of a machine learning origin - is the quality of the forces it predicts relative to an appropriate reference. We choose a random sample of $1.5 \times 10^4$ atomistic reference points from our data and compare the forces as predicted by our model to those from DFT. Additionally, we compare the forces predicted by a number of other popular methods for atomistic modelling; DFT with common exchange correlation functionals, density functional tight binding (DFTB), a number of empirical many body potentials; Tersoff, REBO, AIREBO, AIREBO-Morse and LCBOP, a ReaxFF potential parameterized for condensed carbon and the recently published GAP model for the amorphous phase.[19–22,24,26,44,68] Force errors for the graphene GAP, LCBOP,

|   | 2b | 3b | SOAP |
|---|---|---|---|
| $\delta$(eV) | 10 | 3.7 | 0.07 |
| $r_{\text{cut}}$(Å) | 4.0 | 4.0 | 4.0 |
| $w_{\text{cut}}$(Å) | 1.0 | 1.0 | 1.0 |
| Sparse Method | uniform | uniform | CUR |
| $N_t$ | 50 | 200 | 650 |

TABLE I. Additional parameters used for the training of the GAP model. $\delta$ indicates the relative weighting of the different descriptors, $r_{cut}$ indicates the cutoff width of the descriptor and $w_{cut}$ indicates the characteristic width over which the descriptor magnitude goes to 0. 2b, 3b and MB indicate the two body, three body and many body descriptors used in the construction of the potential. $N_t$ indicates the number of sparse points chosen for each descriptor during training, while the sparse method denotes the method by which sparse points were chosen. More information can be found in the GAP code documentation at [http://www.libatoms.org].

| Potential | RMSE (In-plane) eV Å$^{-1}$ | RMSE (Out-of-plane) eV Å$^{-1}$ | Lattice parameter (0 K) Å | Time (Relative) |
|---|---|---|---|---|
| Graphene GAP | 0.028 | 0.019 | 2.467 (+0.003) | 340 |
| Amorphous GAP | 0.270 | 0.258 | 2.430 (-0.03) | 380 |
| Tersoff | 3.122 | 0.542 | 2.530 (+0.08) | 1 |
| REBO | 0.722 | 0.187 | 2.460 (-0.004) | 1.2 |
| AIREBO | 0.548 | 0.414 | 2.419 (-0.05) | 1.9 |
| AIREBO-Morse | 0.720 | 0.568 | 2.459 (-0.005) | 2.9 |
| LCBOP | 0.595 | 0.306 | 2.459 (-0.005) | 2.3 |
| ReaxFF | 1.226 | 0.311 | 2.462 (-0.002) | 23 |
| DFTB | 0.693 | 0.162 | 2.470 (+0.006) | 950 |
| DFT (optB88-vdW) |  |  | 2.464 | $2 \times 10^7$ (AIMD) |
| Exp.[65] Graphite, 300 K |  |  | 2.462 |  |

TABLE II. Root mean squared force errors, lattice parameters predicted and relative costs of empirical many-body and GAP models. The details for other common DFT functionals tested are available in the SM.

Tersoff and DFTB methods are shown in Fig. 1, where we have separated the data into forces in the 'in-plane' directions and those in the 'out-of-plane' direction. Root mean squared errors (RMSE) are given for all methods in Table II, plots of force correlations and errors for all methods can be found in the SM. We calculate the cost of each of the methods over $10^4$ identical MD steps for 200 atoms, which we normalise for the number of cores on which the simulation was run. Fig. 1 shows that the predictions of the graphene GAP model align very closely with those of the reference DFT method. Forces are obtained with an RMSE of 0.028 eV Å$^{-1}$ in the in-plane direction, and 0.019 eV Å$^{-1}$ in the out of plane direction. The errors obtained from the DFTB and LCBOP methods are much larger, RMS errors in forces are 0.69 and 0.55 eV Å$^{-1}$ respectively and maximum errors of 2 eV Å$^{-1}$ are observed in the worst cases. Errors are largest for the Tersoff potential, for which the RMSE is measured as 3.1 eV Å$^{-1}$ with a maximum in excess of 11 eV Å$^{-1}$. Despite the AIREBO-Morse potential being a more recent iteration of the AIREBO potential (including a Morse potential to model bonding interactions) we find that the modifications are actually a detriment to the quality of the predicted forces, despite the increased cost (Table II).

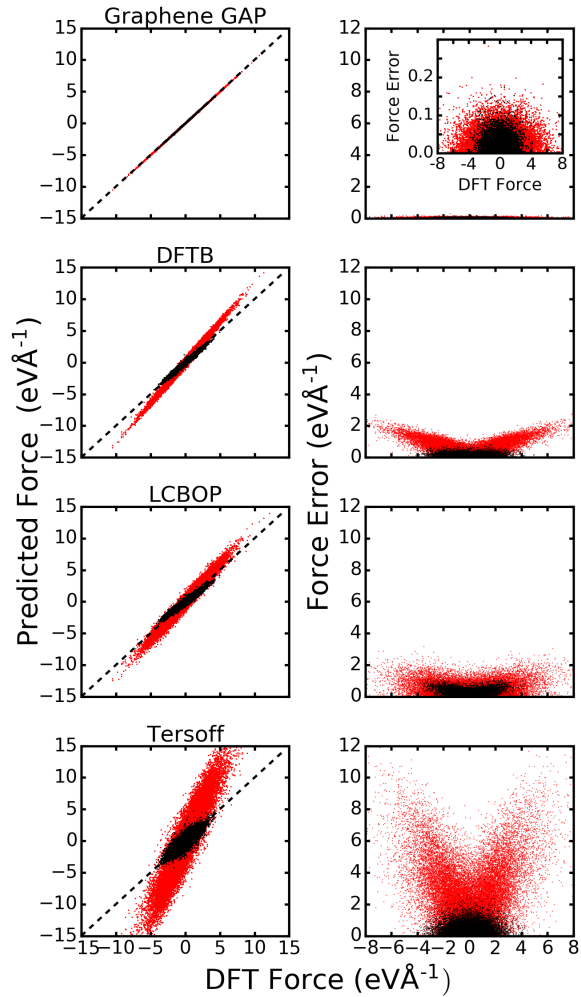

FIG. 1. Force correlations (left) and associated force errors (right) on an independent reference dataset of configurations for the graphene GAP model, DFTB, LCBOP and Tersoff potentials as compared to the reference DFT method, the plots for all methods considered can be found in the SM. Black points indicate forces perpendicular to the plane of the graphene sheet (out-of-plane) while red points indicate forces oriented in the plane. The inset in the graphene GAP plot has a different scale on the y-axis to show more clearly the distribution of force errors, which are smallest for large forces with a Gaussian distribution.

It is important to briefly consider how these conclusions may be affected by the choice of reference method; there are many instances in the literature of disagreement between various exchange correlation functionals and it is important to evaluate the importance of this in the context of graphene, the details of which we give in the SM. We find that there is a minimal dependence of the measured forces on the choice of exchange correlation functional for this system, on average 0.026 eV Å$^{-1}$ in the in-plane and 0.018 eV Å$^{-1}$ in the out of plane direction - indicating that the relative ranking of the benchmarked methods would be the same irrespective of the chosen reference method. Furthermore, the expected performance of the graphene GAP model would also be insensitive to this choice. This is supported by the similarity in the phonon spectra calculated with each of the functionals, which are also available in the SM.

## V. LATTICE PARAMETERS AND IN-PLANE THERMAL EXPANSION

The lattice parameter is a fundamental property for any atomistic model of a material to predict. Many intrinsic properties of materials such as graphene are affected by the lattice constant, while the degree and type interaction between two distinct materials can vary dramatically based on the degree of lattice matching between their two structures.[69] In addition to the ground state lattice parameter, the thermal expansion of graphene is also of interest as it provides insight into the relative strengths of the in-plane and out of plane forces, the anharmonicity of the bonding interactions and the coupling between harmonic and anharmonic vibrational modes.

The nature of the thermal expansion of graphene is, however, a topic wherein many conflicting computational reports may be found.[12,50–52] The experimental coefficient of thermal expansion of freestanding graphene is generally accepted to be negative at moderate temperatures - low lying bending phonon modes cause graphene to 'crumple' and thus shrink in the in-plane direction.[12,50] Graphene has been found from Raman spectroscopy and micromechanical measurements to have a negative in-plane coefficent of thermal expansion at temperatures between 30 and 500 K.[51,52] However, graphene must typically be investigated experimentally while adsorbed on a substrate material, the strain induced from this significantly affects both its 0 K lattice parameter and the thermal expansion of the material, leaving the study of freestanding graphene as a particularly attractive topic for theoreticians.[11,70] *Ab initio* investigations broadly agree in their prediction that the CTE of graphene is negative over a moderate temperature range - but differ in their predictions at higher temperatures. Results from DFPT show non-monotonic behavior, a negative and in-plane coefficient of thermal expansion up to 2000 K, with a minimum at 300 K.[71] Green's function lattice dynamics calculations have found the sign of the CTE to change from negative to positive at temperatures above 500 K and AIMD simulations have found the CTE to be weakly negative over a large temperature range.[11,72] Results from studies employing empirical potentials vary more substantially, the REBO potential predicts a positive CTE over a wide temperature range, the Stillinger-Weber and LBOP potentials predict the CTE to be entirely negative and the LCBOP and LCBOPII[73] potentials predict a change in the sign of the CTE around 500 K.[9,12]

We now compare to lattice parameters over a range of temperatures as predicted by *ab initio* molecular dynamics simulations of graphene sheets using the method established in Ref.[11]. In-plane lattice parameters were averaged over AIMD simulations on freestanding graphene sheets containing 200 atoms between 60 and 2500 K. Calculations were performed at the gamma point, using the optB88-vdW functional and a projector augmented wave potential with a plane wave cutoff of 400 eV, in the NPT ensemble as implemented in VASP, with the constant pressure algorithm applied only in the lateral directions (in-plane).[9,11,74] Three independent simulations at each temperature were conducted and statistics were collected for between 40 and 95 ps depending on the temperature until the lattice parameter was converged to within $10^{-4}$Å. We note that this approach neglects the effect of the zero-point vibrational energy (ZPE) on the calculated lattice parameter and thermal expansion. The inclusion of this has previously been found to increase the ground state lattice parameter of graphene by 0.3%.[71] The effect of ZPE could be included via path-integral type methods, but we consider this unnecessary for the benchmarking purposes of the current study.

Lattice parameters for the empirical and GAP potentials were determined similarly. We performed NPT simulations using the Nosé-Hoover thermostat on freestanding graphene sheets containing 200 atoms. Simulations were equilibrated for 5 ns and statistics collected on three replica simulations over a further 5 ns for each potential, in each case the time averaged lattice parameters were converged to within $10^{-4}$ Å. The coefficient of thermal expansion of graphene is calculated as,

$$\text{CTE} = \frac{1}{A_T} \frac{\partial A_T}{\partial T}. \tag{12}$$

Here, $A$ denotes the area of the graphene sheet and $T$ the temperature in Kelvin. To calculate the CTE we interpolate between calculated data points by fitting splines to the data - we take the derivatives of the fitted splines to evaluate equation 12. The optimized lattice parameters at 0 K for graphene for all methods are also given in Table II for comparison.

The calculated lattice parameters from ground state optimization are given in Table II. The majority of the empirical potentials considered accurately predict the 0 K lattice parameter (with errors typically less than 0.2%), which is found from DFT to be 2.464 Å. The exceptions to this are the Tersoff, AIREBO and Amorphous GAP potentials. The Tersoff potential is found to overestimate the lattice parameter of graphene by 3.2%, while the AIREBO and amorphous carbon potentials underestimate by 2.0% and 1.2% respectively. DFTB would generally be expected to represent an improvement over empirical potentials, however in this instance predicts the lattice parameter of graphene with an error of +0.3%, representing an improvement over only the three worst empirical potentials. The Graphene GAP and ReaxFF potentials are both in excellent agreement with our *ab inito* results with errors of 0.1%.

Most of the potentials considered predict a much larger dependence of the in-plane lattice parameter on the temperature than is calculated from AIMD, which predicts an overall maximum change in value of 0.1% as can be seen from Figure 2B. Our first principles calculations predict a contraction of the graphene sheet up to approximately 1750 K, above which we observe expansion in the in-plane direction. Our graphene GAP model is in excellent agreement with the predictions of the first principles calculations both in terms of the absolute and relative lattice parameters. The relative predictions of the Tersoff potential are also found to be in good agreement with *ab initio* results at low temperatures, despite the significant overestimation of the absolute lattice parameter. The AIREBO and AIREBO-Morse potentials significantly overestimate the in-plane expansion of graphene at moderate temperatures, while the REBO potential predicts an in-plane lattice parameter which increases over the entire observed temperature range. The predictions of the LCBOP potential are in line with those of previous studies, it predicts a strongly negative thermal expansion with a minimum close to 1000 K.[12] The ReaxFF potential considered here is observed to predict a very strong, negative thermal expansion coefficient and predicts the fragmentation of the graphene sheet at temperatures above 1500 K, well below the experimentally determined melting point. Between temperatures of 60 and 1500 K, ReaxFF predicts a strong contraction of the in-plane lattice parameter as a result of large out-of-plane displacements. Figure 2C shows the values for the CTE of graphene as calculated with each of the potentials and with *ab initio* calculations. The LCBOP, AIREBO and AIREBO-Morse potentials predict CTEs which are strongly temperature dependent, switching from negative to positive at temperatures between 500 and 1000 K. The REBO potential similarly predicts a strong temperature dependence, however in this case the CTE is predicted to be positive over the entire measured range. In contrast, the GAP, Tersoff and AIMD simulations predict a much weaker temperature dependence of the CTE, with a change in sign close to 1000 K. The Tersoff potential predicts a continued increase of the in-plane CTE throughout the measured temperature range, while the GAP and AIMD calculations predict a slowdown in the increase and a plateau above 1500 K. Overall it is clear that, the lattice expansion of graphene represents a challenging property to evaluate with molecular dynamics, the GAP model introduced here quantitatively

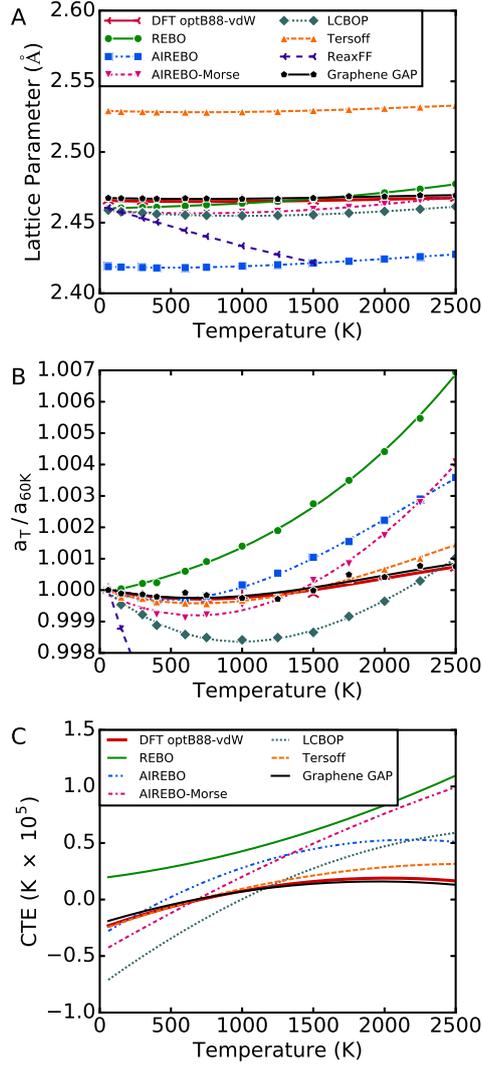

FIG. 2. A) Thermal dependence of the lattice parameter of graphene between 60 and 2500 K, for a range of potentials as compared to the reference value calculated from *ab initio* molecular dynamics calculations. B) Thermal dependence of lattice parameter, a, normalized according to the predicted value at 60 K, emphasising the relative behavior of the different methods - a range of predictions is observed, from monotonically increasing or decreasing lattice parameters to more complex non-monotonic behavior in the case of GAP, LCBOP and AIMD calculations. C) Computed thermal expansion coefficients for graphene as a function of temperature calculated using equation 12, for DFT and various potentials.

reproduces the results of the reference calculations.

## VI. PREDICTION OF PHONON SPECTRA

A correct description of the lattice dynamics of a material is a fundamental requirement for any atomistic model. This experimentally measurable property of a material is obtained computationally directly from the derivative of the forces acting upon the atoms. There is thus a natural and close link between the quality of the phonon spectrum and the quality of the predicted forces with respect to experiment. This makes the prediction of the phonon spectrum an excellent independent metric of the overall quality of a potential. Furthermore, a number of thermodynamic properties of materials, for example the heat capacity, may be obtained directly from dispersion relations via calculation of the free energy.

We use two methods to calculate the phonon spectrum of graphene. To calculate the 0 K phonon spectrum, we use the finite displacement method as implemented in PHON.[75] In order to predict the anharmonic phonon spectrum at finite temperature, we evaluate the elastic constants and thus the phonon spectrum directly from the forces and displacements sampled from MD trajectories.[76,77] As our reference, we compare our results to those determined from the fifth nearest neighbor force constant fit to data measured experimentally using x-ray diffraction (XRD) on graphite.[6,8] The phonon spectrum of graphene is comprised of six branches; ZA, TA, LA, ZO, TO and LO. At the $\Gamma$ point, the LO and TO phonon branches take on the symmetry label $E_{2g}$, the ZO branch is labelled $B_{2g}$ and the lowest energy LA, TA and ZA branches together as $A_{2u}$ and $E_{1u}$.[78] Figure 3 shows the phonon spectra predicted using each of the potentials compared to the reference XRD data. The graphene GAP model achieves excellent agreement with experiment; it correctly predicts the phonon frequencies at almost all of the high symmetry points with sub-meV accuracy. The dispersion behavior of each of the bands is also accurately predicted across all of the sampled regions of the Brillouin zone. The LCBOP and REBO potentials perform comparably to one another, qualitatively correctly predicting the shape and dispersion character of most of the phonon branches. What can be seen in more detail from Figure 4 is that LCBOP achieves a greater accuracy than REBO close to the $\Gamma$ point, but amasses more significant errors overall, on the order of 20 meV, towards the $K$ and $M$ high symmetry points. Conversely, the error in the prediction of the phonon frequencies made by the REBO potential is a much flatter function of k-space with an overall mean absolute error (MAE) of 10 meV. However, both potentials exhibit significant errors in the prediction of the highest energy longitudinal optical (LO) branch, with peak errors of 40 meV and 60 meV for LCBOP and REBO respectively. As would be expected, both the AIREBO and AIREBO-Morse potentials perform comparably, with notable underestimations of the transverse optical (ZO) phonon modes at the $\Gamma$ point. The MAE of each potential is again a relatively flat function of k-space, at 20 meV in both cases. The dispersive character and $B_{2g}$ $\Gamma$ point frequency predicted by DFTB are in good agreement with the experimental results, the most most notable error being the overestimation of the $E_{2g}$ symmetry frequency at the $\Gamma$ point, which is overestimated by 20 meV. We find that the ReaxFF potential provides a reasonably good estimate of dispersion of the low frequency phonon modes, however fails for the highest energy LO and TO branches. This is the case in particular away from the $\Gamma$ point, for which peak errors in the LO branch are found to be in excess of 60 meV. The Tersoff potential, finally, is shown to fail in predicting the energies and dispersion behaviors of all but the two lowest energy branches of the phonon spectrum. Band errors are as large as 110 meV for the $E_{2g}$ symmetry (LO and TO) bands at the $\Gamma$ point, with a MAE across the sampled region of k-space of 40 meV. Although a modified version of the Tersoff potential has been constructed which was optimized to reproduce the lowest energy phonon dispersion modes of graphene, we find that the stability of this potential is not satisfactory due to the reparametrization, and have therefore not included it here.[9,79] We note that an error common to all of the empirical potentials is a failure to describe the dispersive behavior of the high energy LO branch of the phonon spectrum - which the graphene GAP model predicts with negligible error.

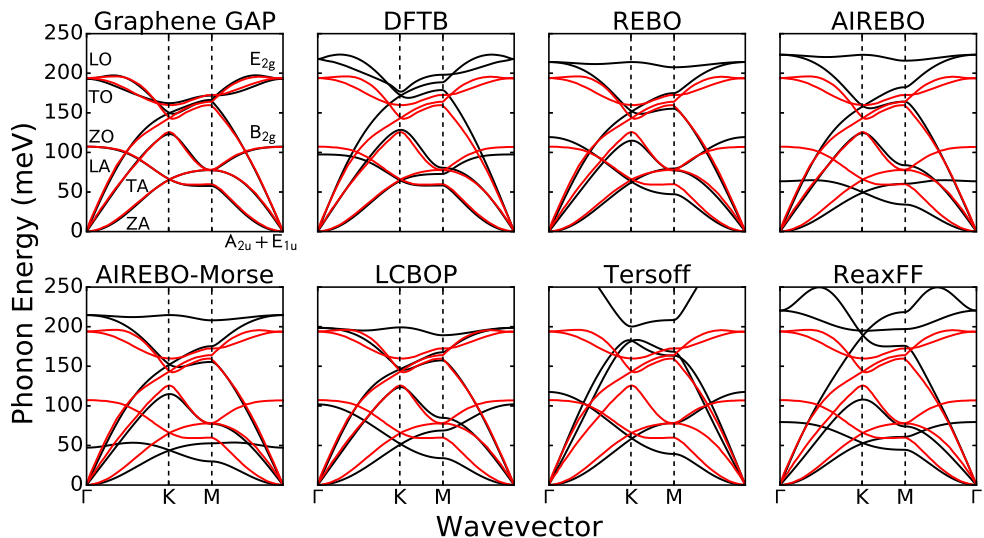

FIG. 3. Comparison of model predictions using the finite displacement method[75] to phonon dispersion from XRD.[76,77] Black lines represent the calculated phonon spectrum, red is the reference XRD. The GAP model accurately reproduces the experimentally determined phonon spectrum over all of the high symmetry directions considered. Labels for branches are shown on the Graphene GAP plot (left) along with symmetry labels at the $\Gamma$ point (right). Note that the highest energy LO branch is not shown for the Tersoff potential in this figure - this branch crosses the $\Gamma$ point at approximately 350 meV.

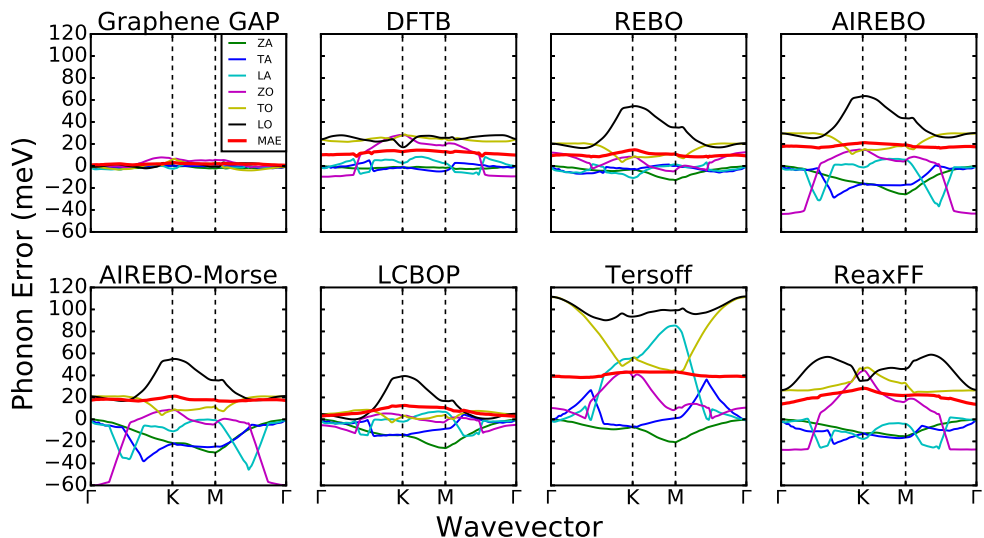

FIG. 4. Absolute errors in prediction of phonon band frequencies along the high symmetry directions in the graphene Brillouin zone, separated by phonon branch type. The thick red line denotes the mean absolute error (MAE) summed across all bands. Notable similarities in the error predicting the character of the LO branch can be seen across the LCBOP, REBO and AIREBO(-Morse) potentials (black line).

In addition to a consideration of the phonon spectrum at a single temperature, we can compare the behavior of particular phonon modes as a function of temperature to experimental observations from Raman spectroscopy. The G band of the graphene phonon spectrum may be unambiguously assigned to the frequency of the $E_{2g}$ symmetry phonon mode at the $\Gamma$ point. We may therefore make a direct comparison between the experimentally measured thermal softening of this mode and the softening predicted by each of the potential models. The correct description of the thermal character of this band is of great importance for the technological application of graphene - the degree of population of the $E_{2g}$ band has implications for the ballistic energy transport which makes graphene so attractive as an electronic material.[50,80] One aspect of this characterization is the correct prediction of the energy of this mode at the $\Gamma$ point, the comparison for which shown in Figure 5 where the phonon spectra for graphene from 60 to 2500 K are given.

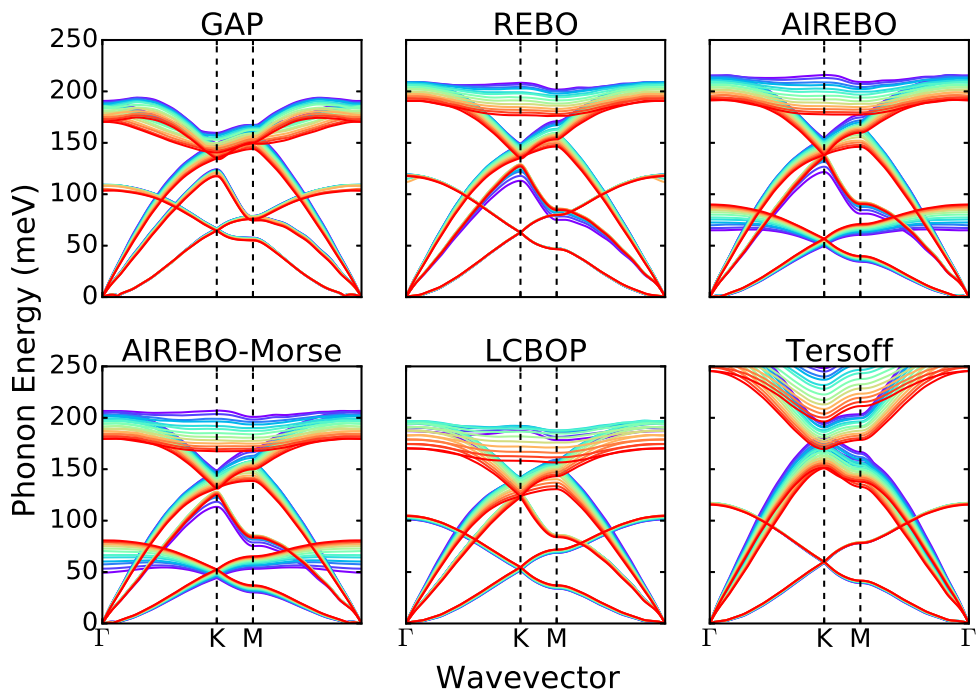

FIG. 5. Finite temperature phonon calculations for graphene simulations between 60 and 2500 K derived directly from molecular dynamics simulations. Strong thermally induced dispersion is seen for the highest energy $E_{2g}$ symmetry phonon modes across all potentials, corresponding to the observed thermally induced dispersion of the Raman G band of graphene. Varying predictions are made for the transverse optical (ZO) branch's dependence on temperature - the AIREBO(-Morse) potentials predict this to have a strong thermal dispersive character. Blue corresponds to simulations at 60 K, through to 2500 K for red in a linear scale.

For each temperature we use the lattice parameter determined for each potential for the given temperature as calculated using the same procedure for determining the lattice parameter described above. Simulations were run for each lattice parameter and each potential in the NVT ensemble using Langevin dynamics. Configurations were first equilibrated for 2 ns until the temperature had equilibrated and statistics were collected over 30 ns trajectories at each temperature, in each case the phonon frequencies of the degenerate LO/TO ($E_{2g}$) branches at the $\Gamma$ point were converged to within 1 meV.

We observe that all potentials predict a large degree of thermally induced dispersion in the highest energy LO/TO branches (Figure 5). The AIREBO and AIREBO-Morse potentials both predict a strong dependence of the trans-verse optical (ZO) branch on temperature, which is not observed for the other methods considered. We compare quantitatively the results of our calculations to those obtained from the variable temperature Raman scattering measurements.[70] The thermally induced dispersion of the Raman G band was measured between 150-900 K for graphene sheets adsorbed on a SiN substrate. The effect of the substrate on the position and thermal dispersion of the G band is two-fold, a constant offset induced by the mismatched lattice parameter and interlayer interactions between the substrate and the graphene and an effect due to the thermally induced strain from the different thermal expansions of the two materials. To account for the first effect, we simply report the change in G band frequency rather than the absolute value. The effect of the differing lattice expansion of the materials may be accounted for by calculating the induced strain and correcting the data using the known biaxial strain coefficient of the graphene G band.[51,70]

$$\Delta\omega_G^s(T) = \beta \int_{T_0}^{T} [\text{CTE}_{\text{sub}}(T) - \text{CTE}_{\text{gr}}(T)]dT \quad (13)$$

Where $\text{CTE}_{\text{sub}}$ and $\text{CTE}_{\text{gr}}$ represent the CTEs of the substrate (SiN) and graphene respectively, and $\beta$ is the known biaxial strain coefficient of graphene ($\beta = -70 \pm 3\,\text{cm}^{-1}/\%$).[81,82] We use values for the CTE graphene as determined by our earlier *ab initio* calculations. Figure 6 shows the thermally induced dispersion of the $E_{2g}$ symmetry phonon modes at the $\Gamma$ point. Our graphene GAP model is seen to be in good agreement with the experimentally observed effects as are the predictions of both the AIREBO and REBO potentials. The AIREBO-Morse potential slightly overestimates the degree of dispersion while the Tersoff potential predicts a significantly enhanced effect. Surprisingly, despite the good predictions of the shape of the phonon dispersion curves by the LCBOP potential using the finite displacement method, we find here a strong qualitative disagreement with the experimental results.

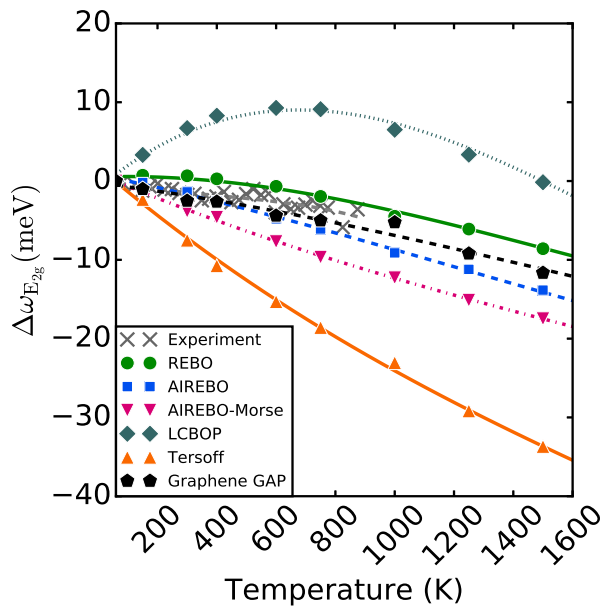

FIG. 6. Change in $\Gamma$ point frequencies for graphene $E_{2g}$ symmetry vibrational mode in the region of 150-1400 K. Compared with results from variable temperature Raman spectroscopy, which have been corrected for the strain induced by the adsorption of the graphene sheet onto the SiN substrate.

## VII. CONCLUSIONS AND DISCUSSION

We have used the Gaussian Approximation Potential method to construct a machine learning potential for graphene, which we have trained using energies, forces and virial stresses calculated using high quality vdW inclusive DFT calculations. We have benchmarked the quality of this potential alongside a number of other commonly used potentials against both *ab initio* and experimental references. We find that the graphene GAP model predicts quantitatively the lattice parameter, coefficient of thermal expansion and phonon properties of graphene. Among the other potentials considered, many of them provide reasonable predictions of one property, but none is successful in predicting the whole range of properties considered. We find the REBO potential to be the best empirical model, providing a good overall description of the lattice dynamics of graphene, including accurately describing the effect of temperature on these. However, despite accurately predicting the 0 K lattice parameter, the REBO potential's predicted dependence of the in-plane lattice parameter is in qualitative disagreement with the results of *ab inito* calculations. In fact, we find that none of the empirical many-body potentials accurately predicts both the 0 K lattice parameter of graphene and the lattice expansion at finite temperature.

The GAP method is computationally more demanding than the empirical many-body potentials considered here, but approximately four orders of magnitude cheaper than direct *ab initio* molecular dynamics, for 200 atoms. Even taking into consideration the computational cost of the generation of the training database, this represents a significant reduction in computational cost with only a marginal compromise on accuracy. Since the scaling of the cost of the GAP model with system size is the same as that of a force-field MD simulation, compared with the $O(N_{\text{electron}}^3)$ scaling of DFT, this reduction in cost would be more effective for larger system sizes. The purpose of the GAP framework is to provide an accuracy close to that of AIMD at a much reduced cost, rather than offering a universally applicable alternative to empirical potentials. Such a potential would be best put to use in cases where a highly accurate description of dynamics is mandated. One such example may be the description of adsorbate diffusion on or confined by graphene sheets, a process which is in some cases strongly enhanced by a coupling between adsorbed molecules and particular graphene phonon modes.[83,84] In this instance, the accurate finite temperature description of the phonon modes provided by the GAP model would be highly desirable. The GAP model would also be ideally suited to modelling thermal transport in graphene nanoelectronic devices, such as transistors. Such systems require highly accurate modelling of heat dissipation, but involve systems of sizes which are beyond the reach of routine *ab initio* calculations.[14–16] In many cases, such as for exotic or newly discovered materials, computational investigations may be hampered by the absence of a well parametrized empirical potential. The GAP framework provides a systematic pathway for the development of specialized potentials in these cases.

Despite the promising behavior of the GAP model considered here, it is important to note that the transferability of the various models may also be an important property. While the GAP model presented here is exemplary in its treatment of free-standing graphene, it is (by construction) not transferable to other phases of carbon i.e. diamond, which the other empirical potentials are capable of. The inability of current machine learning models to extrapolate into foreign regions of chemical space is a well documented one, and great care and attention must be paid to generate a machine learning potential which is capable of treating a wide range of phases of a material.[29,44] Nevertheless, given the systematically improvable nature of Gaussian approximation potentials, a highly accurate and generalized machine learning carbon potential could soon be feasible.

## VIII. ACKNOWLEDGEMENTS

A. M. is supported by the European Research Council under the European Union's Seventh Framework Programme (FP/2007-2013) / ERC Grant Agreement number 616121(HeteroIce project). A.M. is also supported by the Royal Society through a Royal Society Wolfson Research Merit Award. We are grateful to the UK Materials and Molecular Modelling Hub for computational resources, which is partially funded by the EPSRC (EP/P020194/1). We are also grateful for computational support from the UK national high performance computing service, ARCHER, for which access was obtained via the UKCP consortium and funded by EPSRC grant ref EP/P022561/1. In addition, we are grateful for the use of the UCL Grace High Performance Computing Facility (Grace@UCL), and associated support services, in the completion of this work.

[1] Y. Zhang, Y. Tan, H. Stormer, and P. Kim, Nature **438**, 201 (2005).

[2] C. Lee, X. Wei, J. W. Kysar, and J. Hone, Science Magazine **321**, 385 (2008).

[3] A. H. Castro Neto, F. Guinea, N. M. R. Peres, K. S. Novoselov, and A. K. Geim, Reviews of Modern Physics **81**, 109 (2009).

[4] P. Avouris, Z. Chen, and V. Perebeinos, Nature Nanotechnology **2**, 605 (2007).

[5] F. Bonaccorso, Z. Sun, T. Hasan, and A. C. Ferrari, Nature Photonics **4**, 611 (2010).

[6] M. Mohr, J. Maultzsch, E. Dobardžić, S. Reich, I. Milošević, M. Damnjanović, A. Bosak, M. Krisch, and C. Thomsen, Physical Review B **76**, 035439 (2007).

[7] J. A. Yan, W. Y. Ruan, and M. Y. Chou, Physical Review B **77**, 125401 (2008).

[8] J. Maultzsch, S. Reich, C. Thomsen, H. Requardt, and P. Ordejón, Physical Review Letters **92**, 075501 (2004).

[9] Y. Magnin, G. D. Förster, F. Rabilloud, F. Calvo, A. Zappelli, and C. Bichara, Journal of Physics: Condensed Matter **26**, 185401 (2014).

[10] A. Geim, Science **324**, 1530 (2009).

[11] M. Pozzo, D. Alfè, P. Lacovig, P. Hofmann, S. Lizzit, and A. Baraldi, Physical Review Letters **106**, 135501 (2011).

[12] K. V. Zakharchenko, M. I. Katsnelson, and A. Fasolino, Physical Review Letters **102**, 046808 (2009).

[13] M. Ma, G. Tocci, A. Michaelides, and G. Aeppli, Nature Materials **15**, 66 (2016).

[14] A. Balandin, Nature Materials **10**, 569 (2011).

[15] V. Varshney, S. S. Patnaik, A. K. Roy, G. Froudakis, and B. L. Farmer, ACS Nano **4**, 1153 (2010).

[16] F. Schwierz, Nature Nanotechnology **5**, 487 (2010).

[17] C. P. Herrero and R. Ramírez, Physical Review B **79**, 115429 (2009).

[18] C. P. Herrero and R. Ramírez, Journal of Chemical Physics **145**, 224701 (2016).

[19] J. Tersoff, Physical Review Letters **61**, 2879 (1988).

[20] J. Tersoff, Physical Review B **39**, 5566 (1989).

[21] D. W. Brenner, Physical Review B **42**, 9458 (1990).

[22] S. J. Stuart, A. B. Tutein, and J. A. Harrison, The Journal of Chemical Physics **112**, 6472 (2000).

[23] T. C. O'Connor, J. Andzelm, and M. O. Robbins, Journal of Chemical Physics **142**, 024903 (2015).

[24] J. H. Los and a. Fasolino, Physical Review B **68**, 24107 (2003).

[25] A. C. T. van Duin, S. Dasgupta, F. Lorant, and W. A. Goddard III, Journal of Physical Chemistry A **105**, 9396 (2001).

[26] S. Goverapet Srinivasan, A. C. T. Van Duin, and P. Ganesh, Journal of Physical Chemistry A **119**, 571 (2015).

[27] D. Porezag, T. Frauenheim, T. Köhler, G. Seifert, and R. Kaschner, Physical Review B **51**, 12947 (1995).

[28] G. Seifert, D. Porezag, and T. Frauenheim, International Journal of Quantum Chemistry **58**, 185 (1996).

[29] V. L. Deringer, G. Csányi, and D. M. Proserpio, ChemPhysChem **18**, 873 (2017).

[30] M. Rupp, M. R. Bauer, R. Wilcken, A. Lange, M. Reutlinger, F. M. Boeckler, and G. Schneider, PLoS Computational Biology **10**, e1003400 (2014).

[31] G. Montavon, M. Rupp, V. Gobre, A. Vazquez-Mayagoitia, K. Hansen, A. Tkatchenko, K. R. Müller, and O. Anatole Von Lilienfeld, New Journal of Physics **15**, 095003 (2013).

[32] K. Hansen, G. Montavon, F. Biegler, S. Fazli, M. Rupp, M. Scheffler, O. A. Von Lilienfeld, A. Tkatchenko, and K. R. Müller, Journal of Chemical Theory and Computation **9**, 3404 (2013).

[33] M. Rupp, R. Ramakrishnan, and O. A. von Lilienfeld, Journal of Physical Chemistry Letters **6**, 3309 (2015).

[34] A. Lopez-Bezanilla and O. A. Von Lilienfeld, Physical Review B **89**, 235411 (2014).

[35] K. Hansen, F. Biegler, R. Ramakrishnan, W. Pronobis, O. A. Von Lilienfeld, K. R. Müller, and A. Tkatchenko, Journal of Physical Chemistry Letters **6**, 2326 (2015).

[36] J. C. Snyder, M. Rupp, K. Hansen, K. R. Müller, and K. Burke, Physical Review Letters **108**, 253002 (2012).

[37] L. Li, J. C. Snyder, I. M. Pelaschier, J. Huang, U. N. Niranjan, P. Duncan, M. Rupp, K. R. Müller, and K. Burke, International Journal of Quantum Chemistry **116**, 819 (2016).

[38] M. Rupp, International Journal of Quantum Chemistry **115**, 1058 (2015).

[39] K. Vu, J. C. Snyder, L. Li, M. Rupp, B. F. Chen, T. Khelif, K. R. Müller, and K. Burke, International Journal of Quantum Chemistry **115**, 1115 (2015).

[40] V. Krková, Neural Networks **5**, 501 (1992).

[41] J. Behler and M. Parrinello, Physical Review Letters **98**, 146401 (2007).

[42] R. Z. Khaliullin, H. Eshet, T. D. Kühne, J. Behler, and M. Parrinello, Physical Review B **81**, 100103 (2010).

[43] R. Z. Khaliullin, H. Eshet, T. D. Kühne, J. Behler, and M. Parrinello, Nature Materials **10**, 693 (2011).

[44] V. L. Deringer and G. Csányi, Physical Review B **95**, 094203 (2017).

[45] A. P. Bartók, M. C. Payne, R. Kondor, and G. Csányi, Physical Review Letters **104**, 136403 (2010).

[46] J. Behler, International Journal of Quantum Chemistry **115**, 1032 (2015).

[47] J. Behler, Journal of Chemical Physics **145**, 170901 (2016).

[48] A. P. Bartók, R. Kondor, and G. Csányi, Physical Review B **87**, 184115 (2013).

[49] A. P. Bartók and G. Csányi, International Journal of Quantum Chemistry **115**, 1051 (2015).

[50] N. Bonini, M. Lazzeri, N. Marzari, and F. Mauri, Physical Review Letters **99**, 176802 (2007).

[51] D. Yoon, Y. W. Son, and H. Cheong, Nano Letters **11**, 3227 (2011).

[52] V. Singh, S. Sengupta, H. S. Solanki, R. Dhall, A. Allain, S. Dhara, P. Pant, and M. M. Deshmukh, Nanotechnology **21**, 165204 (2010).

[53] S. De, A. P. Bartók, G. Csányi, and M. Ceriotti, Physical Chemistry Chemical Physics **18**, 13754 (2016).

[54] J. Behler, Journal of Chemical Physics **134**, 074106 (2011).

[55] O. A. Von Lilienfeld, R. Ramakrishnan, M. Rupp, and A. Knoll, International Journal of Quantum Chemistry **115**, 1084 (2015).

[56] G. Kresse and J. Hafner, Physical Review B **47**, 558 (1993).

[57] G. Kresse and J. Furthmüller, Computational Materials Science **6**, 15 (1996).

[58] G. Kresse and J. Furthmüller, Physical Review B **54**, 11169 (1996).

[59] M. Dion, H. Rydberg, E. Schröder, D. C. Langreth, and B. I. Lundqvist, Physical Review Letters **92**, 246401 (2004).

[60] J. Klimeš, D. R. Bowler, and A. Michaelides, Journal of Physics: Condensed Matter **22**, 022201 (2010).

[61] G. Kresse, Physical Review B **59**, 1758 (1999).

[62] G. Román-Pérez and J. M. Soler, Physical Review Letters **103**, 096102 (2009).

[63] J. Klimeš, D. R. Bowler, and A. Michaelides, Physical Review B **83**, 195131 (2011).

[64] H. Monkhorst and J. Pack, Physical Review B **13**, 5188 (1976).

[65] G. Graziano, J. Klimeš, F. Fernandez-Alonso, and A. Michaelides, Journal of Physics: Condensed Matter **24**, 424216 (2012).

[66] See Supplemental Material at [link placeholder text] for details of functional dependence of forces and phonon dispersion curves and extended figures of empirical potential force errors.

[67] S. Plimpton, Journal of Computational Physics **117**, 1 (1995).

[68] J. H. Los, L. M. Ghiringhelli, E. J. Meijer, and A. Fasolino, Physical Review B **73**, 229901 (2006).

[69] M. Fitzner, G. C. Sosso, S. J. Cox, and A. Michaelides, Journal of the American Chemical Society **137**, 13658 (2015).

[70] S. Linas, Y. Magnin, B. Poinsot, O. Boisron, G. D. Förster, V. Martinez, R. Fulcrand, F. Tournus, V. Dupuis, F. Rabilloud, L. Bardotti, Z. Han, D. Kalita, V. Bouchiat, and F. Calvo, Physical Review B **91**, 075426 (2015).

[71] N. Mounet and N. Marzari, Physical Review B **71**, 205214 (2005).

[72] J. W. Jiang, J. S. Wang, and B. Li, Physical Review B **80**, 205429 (2009).

[73] L. M. Ghiringhelli, J. H. Los, A. Fasolino, and E. J. Meijer, Physical Review B **72**, 214103 (2005).

[74] E. R. Hernández, A. Rodriguez-Prieto, A. Bergara, and D. Alfè, Physical Review Letters **104**, 185701 (2010).

[75] D. Alfè, Computer Physics Communications **180**, 2622 (2009).

[76] L. T. Kong, Computer Physics Communications **182**, 2201 (2011).

[77] C. Campañá and M. H. Müser, Physical Review B **74**, 075420 (2006).

[78] The label 'Z' denotes an out-of-plane vibration, 'L' a longitudinal, in-plane vibration and 'T' a transverse shear mode. Each of these modes may be either acoustic or optical in nature, indicating the phase of the displacements of adjacent nuclei relative to one another. Acoustic phonons represent in-phase vibrational modes, while an optical phonon represents an out-of-phase normal mode of vibration, wherein any two atoms are seen to move against each other.

[79] L. Lindsay and D. A. Broido, Physical Review B **81**, 205441 (2010).

[80] M. J. Biercuk, S. Ilani, C. M. Marcus, and P. L. Mceuen, Physical Review Letters **84**, 2941 (2000).

[81] T. M. G. Mohiuddin, A. Lombardo, R. R. Nair, A. Bonetti, G. Savini, R. Jalil, N. Bonini, D. M. Basko, C. Galiotis, N. Marzari, K. S. Novoselov, A. K. Geim, and A. C. Ferrari, Physical Review B **79**, 205433 (2009).

[82] Z. H. Ni, T. Yu, Y. H. Lu, Y. Y. Wang, Y. P. Feng, and Z. X. Shen, ACS Nano **2**, 2301 (2008).

[83] M. Ma, G. Tocci, A. Michaelides, and G. Aeppli, Nature Materials **15**, 66 (2016).

[84] R. Mirzayev, K. Mustonen, M. R. A. Monazam, A. Mittelberger, T. J. Pennycook, C. Mangler, T. Susi, J. Kotakoski, and J. C. Meyer, Science Advances **3**, e1700176 (2017).

# Supplementary Information for "A Machine Learning Interatomic Potential for Graphene"

Patrick Rowe

*Thomas Young Centre, London Centre for Nanotechnology, and Department of Physics and Astronomy,*
*University College London, Gower Street, London, WC1H 6BT, U.K*

Gábor Csányi

*Engineering Laboratory, University of Cambridge, Trumpington Street, Cambridge CB2 1PZ, U.K.*

Dario Alfè

*Thomas Young Centre, London Centre for Nanotechnology and Department of Earth Sciences,*
*University College London, 1719 Gordon Street, London,*
*WC1H 0AH, U.K. Gower Street, London WC1E 6BT, U.K.*

Angelos Michaelides

*Thomas Young Centre, London Centre for Nanotechnology, and Department of Physics and Astronomy,*
*University College London, Gower Street, London, WC1H 6BT, U.K.*

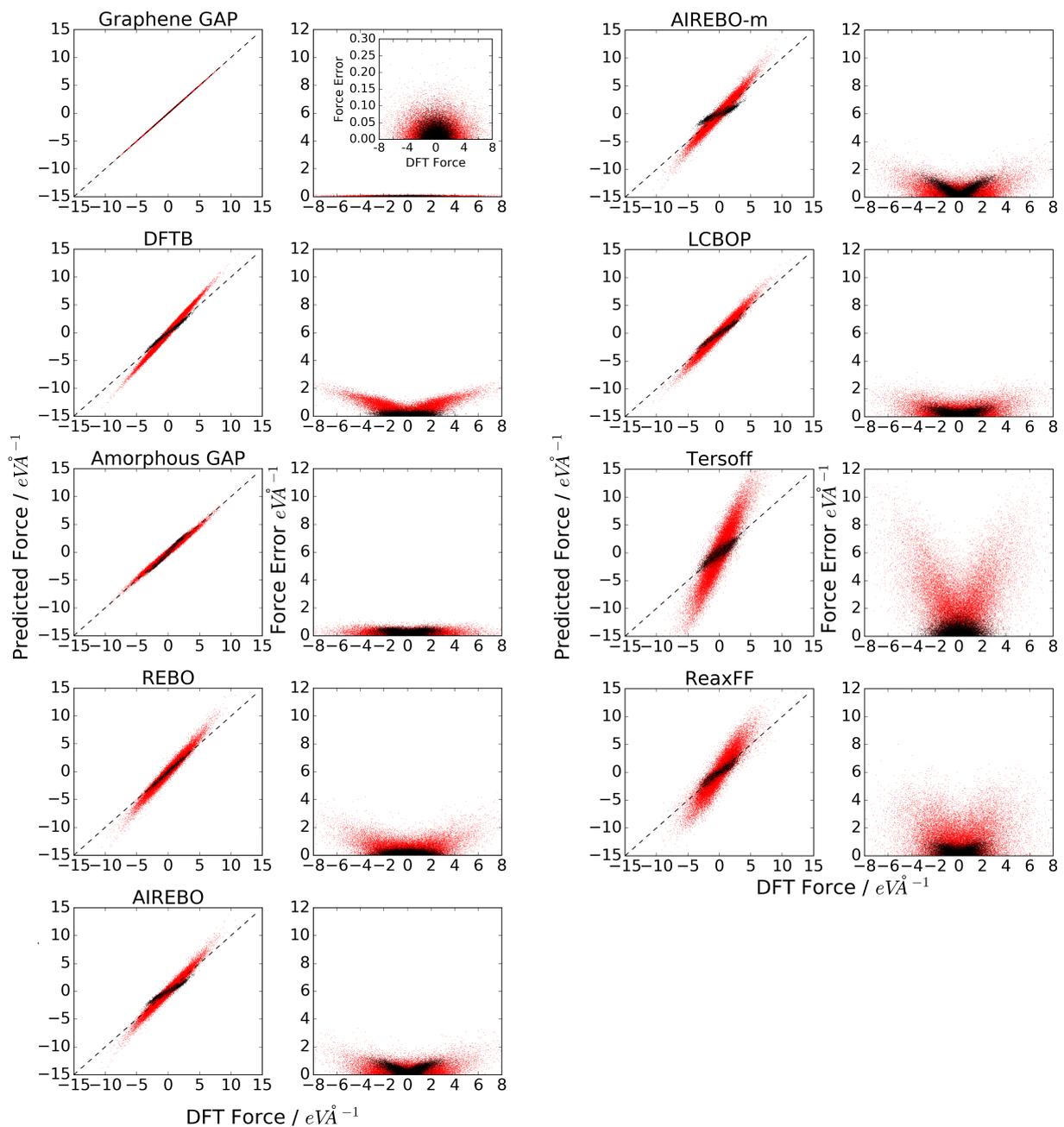

FIG. 1. Force errors for all tested classical potentials, Amorphous and Graphene GAP potentials and DFTB, compared to optB88-vdW dft.

The reference DFT method (optB88-vdW) overestimates the lattice parameter by 0.002 Å (0.08%) with respect to the experimentally determined graphite lattice parameter of 2.462 Å.[3] The remaining DFT functionals considered also overestimate the lattice parameter by between 0.002 and 0.008 Å, with the exception of LDA which underestimates the lattice parameter by 0.016 Å (0.65%) with respect to the experimentally determined value. We note that the force errors associated with the quality of the fit of the GAP model to the DFT reference is in general smaller than or comparable to the force error if measured between two different exchange-correlation functionals.

| Potential | RMSE (In-plane) eV Å$^{-1}$ | RMSE (Out-of-plane) eV Å$^{-1}$ | Lattice parameter (0 K) | Time (Relative) |
| --- | --- | --- | --- | --- |
| Graphene GAP | 0.028 | 0.019 | 2.467 (+0.003) | 344 |
| Amorphous GAP | 0.270 | 0.258 | 2.430 (-0.03) | - |
| Tersoff | 3.122 | 0.542 | 2.530 (+0.08) | 1 |
| REBO | 0.722 | 0.187 | 2.460 (-0.004) | 1.2 |
| AIREBO | 0.548 | 0.414 | 2.419 (-0.05) | 1.9 |
| AIREBO-m | 0.720 | 0.568 | 2.459 (-0.005) | 2.9 |
| LCBOP | 0.595 | 0.306 | 2.459 (-0.005) | 2.3 |
| ReaxFF | 1.226 | 0.311 | 2.462 (-0.002) | 23 |
| DFTB | 0.693 | 0.162 | 2.470 (+0.006) | 950 |
| DFT (optB88-vdW) | | | 2.464 | $2 \times 10^7$ (AIMD) |
| DFT (LDA) | 0.015 | 0.058 | 2.446 (-0.018) | - |
| DFT (PBE) | 0.032 | 0.008 | 2.467 (+0.003) | - |
| DFT (optB86b-vdW) | 0.027 | 0.010 | 2.466 (+0.002) | - |
| DFT (optPBE-vdW) | 0.017 | 0.017 | 2.471 (+0.006) | - |
| DFT (PBE-D3)[1] | 0.032 | 0.008 | 2.467 (+0.003) | - |
| DFT (PBE-TS)[2] | 0.031 | 0.008 | 2.464 (+0.0) | - |
| Exp. (Graphite, 300 K) | | | 2.462 | |

TABLE I. Root mean squared force errors, lattice parameters predicted and relative costs of empirical many-body and GAP models. Bracketed values represent the difference in lattice parameter associated with choosing a different exchange correlation functional rather than optB88-vdW as the reference DFT method for fitting and benchmarking. Dashed entries for timings have not been measured.

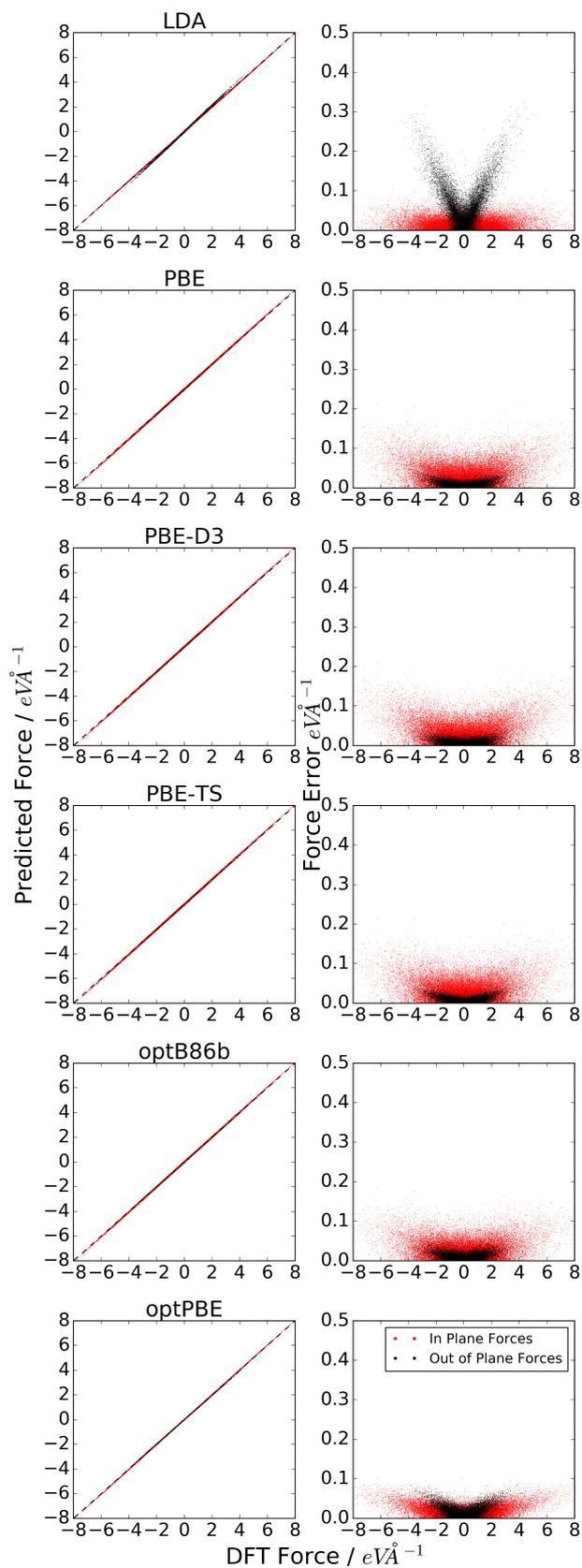

FIG. 2. Force errors for other choices of DFT functional versus the chosen optB88-vdW reference.

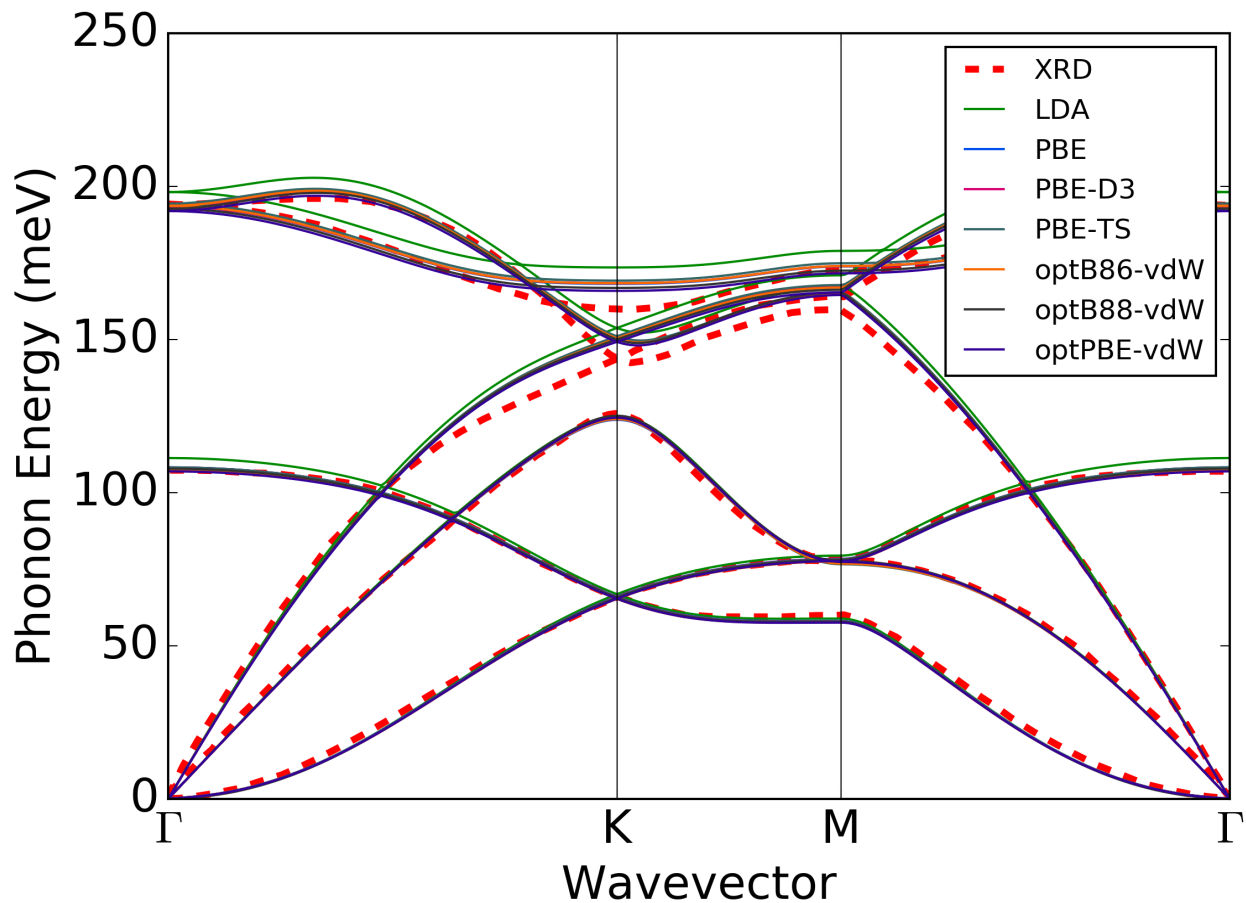

FIG. 3. Comparison of phonon spectra as calculated using the finite displacement method for a variety of DFT functionals, highlighting the robustness of these results to this choice. We note that the errors present in the graphene GAP versus experiment in the main text are the result of a close fit to the reference DFT method - the same inaccuracies close to the K high symmetry point are observed both for the DFT reference and the graphene GAP.

[1] S. Grimme, J. Antony, S. Ehrlich, and H. Krieg, Journal of Chemical Physics **132**, 154104 (2010).

[2] S. Grimme, Journal of Computational Chemistry **27**, 1787 (2009).

[3] G. Graziano, J. Klimeš, F. Fernandez-Alonso, and A. Michaelides, Journal of Physics: Condensed Matter **24**, 424216 (2012).